# Systematic Review of Methods and Prognostic Value of Mitotic Activity – Part 2: Canine Tumors

Christof A. Bertram [1], Taryn A. Donovan [2], Alexander Bartel [3]

[1] University of Veterinary Medicine Vienna, Vienna, Austria

[2] The Schwarzman Animal Medical Center, New York, USA

[3] Freie Universität Berlin, Berlin, Germany

Corresponding author:

Alexander Bartel, Department of Veterinary Medicine, Institute for Veterinary Epidemiology and Biostatistics, Freie Universität Berlin, Koenigsweg 67, Berlin, 14163 Berlin, Germany. Email: Alexander.Bartel@fu-berlin.de

## Abstract

One of the most relevant prognostication tests for tumors is cellular proliferation, which is most commonly measured by the mitotic activity in routine tumor sections. The goal of this systematic review is to scholarly analyze the methods and prognostic relevance of histologically measuring mitotic activity in canine tumors. A total of 137 articles that correlated the mitotic activity in canine tumors with patient outcome were identified through a systematic (PubMed and Scopus) and manual (Google Scholar) literature search and eligibility screening process. These studies determined the mitotic count (MC, number of mitotic figures per tumor area) in 126 instances, presumably the mitotic count (method not specified) in 6 instances and the mitotic index (MI, proportion of mitotic figures per tumor cells) in 5 instances. A particularly high risk of bias was identified in the available details of the MC methods and statistical analysis, which often didn't quantify the prognostic discriminative ability of the MC and only reported p-values. A significant association of the MC with survival was found in 72/109 (66%) studies. However, survival was evaluated by at least three studies in only 7 tumor types/groups, of which a prognostic relevance is apparent for mast cell tumors of the skin, cutaneous melanoma and soft tissue sarcoma of the skin. None of the studies on the MI found a prognostic relevance. Further studies on the MC and MI with standardized methods are needed to prove the prognostic benefit of this test for further tumor types.

## Introduction

Dogs with malignant tumors exhibit a variable clinical course based on certain tumor and patient characteristics.[91] One of the most relevant tumor characteristics regarding prognostication is cellular proliferation.[42,91] While most measurement methods of tumor proliferation require immunohistochemistry (such as the Ki67 index) or special stains (such as the AgNOR score), the most practical approach is to measure mitotic activity in routine Hematoxylin and Eosin-stained tumor sections.[138]

The histological measurement methods for mitotic activity (quantification of mitotic figures, i.e. cells in the M-phase of cell division with histologically distinct features) varies largely in previous studies.[91] Two broad categories can be distinguished: 1) The mitotic count (MC) represents the absolute number of mitotic figures per tumor area and 2) the mitotic index (MI) represents the proportion of mitotic figures among all tumor cells per tumor area.[92] While there are recent efforts, starting in 2016, to standardize the measurement method of the MC,[42,91,92] many studies were published before those guidelines. An overview of the previously applied methods is needed to better understand current practice and to direct future recommendations.

A vast number of studies have evaluated mitotic activity (mostly the MC) as a prognostic test in several canine tumor types (see result section). While mitotic activity is generally considered to be associated with the biological behavior of tumors and outcome of tumor patients,[42,91] there are currently no tumor type-specific recommendations for the routine assessment of mitotic activity. Due to the methodological differences in prognostic studies and the intrinsic bias of observational studies, validation of research findings and summaries through systematic review (and ideally meta-analysis) are needed for each tumor type.[17,90]

The goal of this systematic review is to scholarly analyze the methods and prognostic relevance of histologic measuring mitotic activity in canine tumors. We intend to give an overview of current literature and derive recommendations for routine diagnostic practice and future research goals.

## Material and Methods

This systematic review was conducted in the similar way as a previous systematic review on mitotic activity in feline tumors using the same literature search protocol (with modified search terms and a single literature reviewer), data extraction and article evaluation (risk of bias) criteria.[14]

### Literature search

References were identified through systematic (predefined search terms) and manual (free search terms) search to ensure literature saturation (Fig. 1). Eligibility screening was carried out by a single author (CAB) using a two-step procedure.

Systematic literature identification was carried out in two databases, namely PubMed (1950 to present) and Scopus (1970 to present), on April 30th 2022 using following predefined search terms: (Dog OR dogs OR canine) AND (mitotic count OR mitotic index). Duplicates were removed, and the two-step eligibility screening was carried out in Rayyan [106] using the inclusion/exclusion criteria as provided in Table 1.

Manual literature search in Google Scholar and the perusal of cited references ("cited by" search in google scholar) was conducted in 2022 until 30th April 2022. During literature search, the aforementioned eligibility criteria were applied and only articles that meet these criteria were included. Duplicates to the systematic literature search were excluded.

**Table 1.** Summary of the inclusion/exclusion criteria for the two eligibility screening steps applied to the identified references.

| Screening Step | Decision category | | Inclusion criteria | Exclusion criteria |
|---|---|---|---|---|
| Title-abstract | 1) Study design | | Original study, peer-reviewed | Case-reports, reviews |
| | 2) Topic | a) Species | Dog / Canine | Other species |
| | | b) Tumor | Spontaneous tumors | Experimentally induced tumors |
| | | | Malignant tumors with potential for metastasis | Benign tumors |
| | | c) Prognostic test | Mitotic count (MC), mitotic index (MI) | No mitotic activity measurement |
| | | d) Examination method | Histology | Cytology |
| | 3) Language of main text | | English or German | Other language |
| Full text | 1) Article accessibility | | Article accessible | Article inaccessible |
| | 2) Topic | a) Patient outcome | Correlation of the MC/MI with survival, tumor progression, metastasis, or recurrence | No correlation of the MC/MI with patient follow-up |

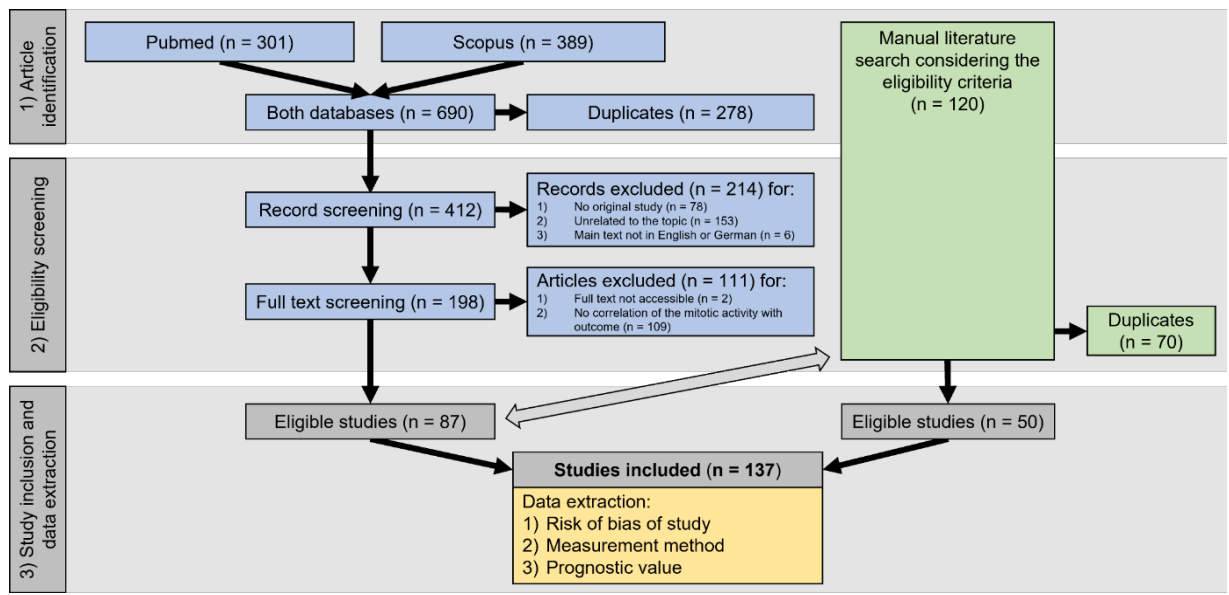

**Figure 1.** Flow diagram (based on the PRISMA recommendation) of the literature search divided into systematic (blue boxes) and manual (green boxes) article identification and eligibility screening, followed by study inclusion with subsequent data extraction.

### Data extraction and analysis

Information regarding the publication (paper identification, year of publication, journal), tumor type evaluated, measurement methods of the MC or MI, prognostic value of mitotic activity was extracted from each article in the same way as previously described.[14] Risk of bias of each study was evaluated (low, moderate, high) specifically for the information regarding the mitotic activity using a preciously developed protocol [14] for more objectivity. The overall risk of bias was based on four domains: 1) study population, 2) outcome assessment, 3) Mitotic activity methods, 4) data analysis.

## Results

### Study selection

The article identification and eligibility screening process are summarized in Fig. 1. Through the systematic literature search, 87 eligible articles out of 412 unique references were identified. Fifty additional articles were found during manual literature search, summing up to a total of 137 articles evaluated in this systematic review.[1-3,5,7-11,16,18-20,22-28,30-36,38-41,43-48,50-63,66-72,75-81,83-89,93-105,107-119,121-137,139-147,149-161]

### Study characterization

All of the included articles were written in English. One third (42/137, 31%) of the publications were published in journals focused on veterinary pathology (Vet Pathol, J Comp Pathol and J Vet Diagn Invest).

Based on the described method, 126 articles evaluated the MC (number of mitotic figures per tumor area),[2,3,5,7-11,16,18-20,22-28,30-36,38-41,43-48,50-63,66,67,69-72,75-81,83-89,93,96-105,110-119,121-123,125-128,130-133,135-137,139-147,149-158,160,161], 5 evaluated the MI (proportion of mitotic figures per number of tumor cells) [95,109,124,134,159] and 6 did not specify the mitotic activity method in their paper.[1,68,94,107,108,129] We assume that the articles without method specification performed the MC, summing up to a total of 132 articles (96.4%). The number of articles included in this review published per year increased over time with more than 10 articles per year between 2018-2021 (Fig. 2).

Whereas the 5 references that determined the MI always used the preferred term "mitotic index", the other 132 articles on the MC use various and sometimes multiple terms including mitotic index (N = 83), mitotic count (N = 40), mitotic rate (N = 13), number of mitosis (N = 3), number of mitotic figures (N = 1), number of mitotic cells (N

= 1), mitotic figures (N = 1), and mitosis (N = 1). Since publishing recommendations on the terminology in 2016 [92], the frequency of the use of correct terminology for MC has improved. While only 12/82 (15%) of the articles published before 2017 used the correct term, 28/55 (51%) after 2016 used the recommended terminology (Fig. 2). Usage of the correct term after 2016 was even higher when the journal had a veterinary pathology focus (12/14, 86%) compared to other journals (16/41, 39%).

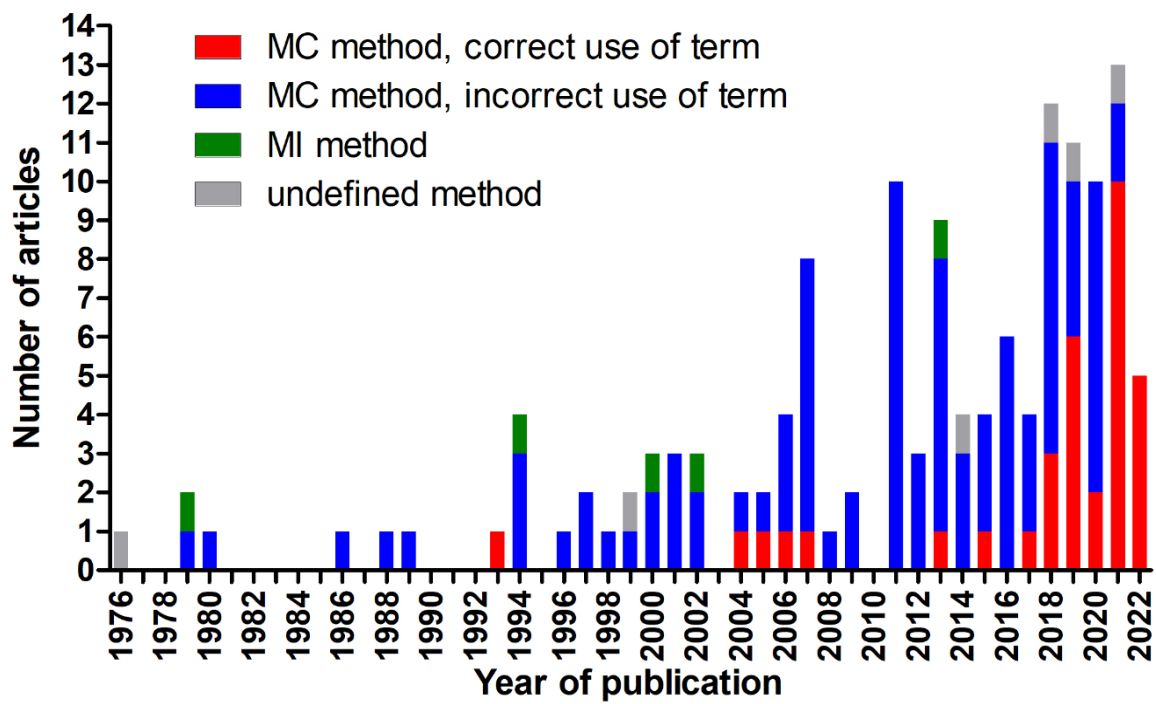

**Figure 2.** Stacked bar chart of the number of publications included in this systematic review per year of publication. The year 2022 includes publications until April 30th.

MC, mitotic count; MI, mitotic index

## Descriptive summary

### Mitotic count (MC)

The 132 studies on the MC evaluated numerous tumor types, whereas some studies included several tumor types or tumor locations or lacked relevant information on these tumor specifications (Table 2).

**Table 2.** Number of articles per tumor type/group and tumor specifications based on tumor types or locations.

| Tumor type/group | Number of references | Tumor specifications of the articles |
|---|---|---|
| Mast cell tumors | 29 (22%) | Skin (N = 24), cutaneous and subcutaneous), skin and mucocutaneous (N = 2), intramuscular (N = 1), oral mucosa (N = 1), unspecified location (N = 1) |
| Soft tissue tumors | 21 (16%) | Skin (N = 13), gastrointestinal (N = 4), visceral (N = 1), smooth muscle tumor (N = 1), appendicular, axial skeleton soft tissue and visceral (N = 1), unspecified location (N = 1) |
| Melanocytic tumors | 20 (15%) | Oral (N = 9), cutaneous (N = 4), oral and cutaneous (N = 5), (intra)ocular (N = 2) |
| Mammary tumors | 14 (11%) | Malignant tumors (N = 8), any (N = 4), carcinoma (N = 1), neuroendocrine carcinoma (N = 1) |
| Osteosarcoma | 9 (7%) | Appendicular (N = 5), mandibular (N = 1), surface (N = 1), any (N = 2) |
| Lymphoma | 8 (6%) | Multicentric (N = 2), diffuse large B-cell (N = 1), diffuse small B-cell (N = 1), Burkitt-like (N = 1), indolent (N = 1), small intestinal (N = 1), any (1) |
| Hemangiosarcoma | 7 (5%) | Splenic (N = 2), cutaneous (N = 1), subcutaneous and intramuscular (N = 1), other than skin (N = 1), non-visceral (N = 1), falciform fat (N = 1) |
| Apocrine gland anal sac adenocarcinoma | 5 (4%) | – |
| Splenic tumors | 4 (3%) | Mesenchymal/stromal sarcoma (N = 2), fibrohistocytic nodules (N = 2) |
| Pulmonary tumors | 3 (2%) | – |
| Renal cell carcinoma | 2 | – |
| Insulinoma | 2 | – |
| Squamous cell carcinoma | 2 | Skin (N = 2) |
| Salivary gland tumors | 1 | – |
| Glial tumors | 1 | – |
| Synovial sarcoma | 1 | – |
| Thymic tumors | 1 | – |
| Pheochromocytoma | 1 | – |
| Esophageal sarcoma | 1 | – |

### Risk of bias

The risk of bias of each of the four domains and the overall risk of bias is summarized in Table 3. Most studies (70/132, 53%) had a high overall risk of bias based on a high risk in at least one of the four domains. One study included some feline cases in the study population,[128] which poses a high risk of bias as equality in the extent of the association of the MC with outcome between different species should not be expected. Data analysis (domain 4) was often restricted to a statistical test of significance and in many studies with non-significant results the actual p-values were not even reported.

**Table 3.** Summary of the risk of bias evaluation for all studies on the mitotic count combined. The overall risk of bias was based on four domains (D1-D4).

| Risk of bias | Number and percent of articles | | | | |
|---|---|---|---|---|---|
| | D1: Study population | D2: Outcome assessment | D3: MC methods | D4: Data analysis | Overall |
| Low ⊕ | 28 (21%) | 17 (13%) | 19 (14%) | 6 (5%) | 4 (3%) |
| Moderate ○ | 68 (52%) | 92 (70%) | 20 (15%) | 61 (46%) | 58 (44%) |
| High ⊖ | 36 (27%) | 23 (17%) | 93 (70%) | 65 (49%) | 70 (53%) |

### Methods

The MC values were taken from pathology records in 19/132 studies (14%), partially taken from pathology reports and newly determined in 2/132 studies (2%) and likely newly determined for the study following the study protocol in the remaining 111 studies (84%). One study employed a semi-quantitative scoring to assessing mitotic density [89], while the rest enumerated mitotic figures. Special staining methods were used in one study (Toluidine blue).[112] Seven of the 20 studies (35%) on melanoma specified that they used bleached slides (for all cases or for heavily pigmented cases), and two studies (10%) assigned a mitotic count of 0 when nuclei were obscured by pigmentation. One study evaluated the MC in digital whole slide images (20x and 40x scan magnification) and glass slides,[154] while the remaining studies (presumably) used light microscopy. The use of automated image analysis was not reported in any study.

A summary of the key methodical aspects of the mitotic count applied in the 132 studies is depicted in Fig. 3. Generally, the proportion of studies that reported the details on the MC methods increased for studies published after 2017 and in journals with a focus on pathology.

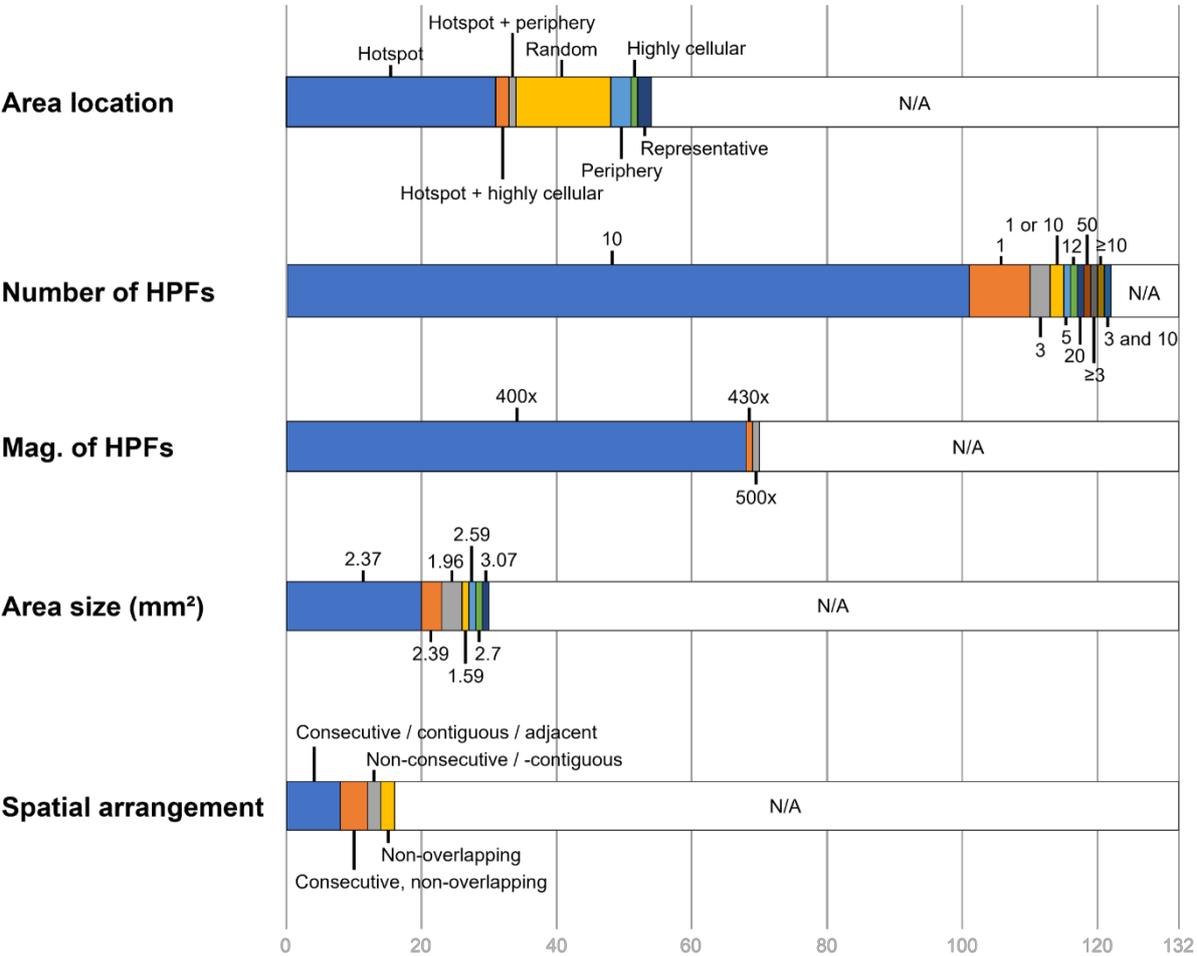

**Figure 3.** Stacked bar chart for the key methodical aspects of the mitotic count applied in the 132 studies.

HPFs, high-power fields; Mag., magnification; N/A, not available

**Prognostic value**

The outcome metrics evaluated in the 132 studies were survival (N = 113, 86%), disease progression (occurrence of metastasis or local recurrence; N = 40, 30%), metastasis (N = 26, 20%), recurrence (N = 22, 17%, particularly for soft tissue tumors), and recurrence of hypoglycemia in insulinoma (N = 2, 2%). The number of complete cases varied between 6 and 384 (median: 50 cases; mean: 64 cases). The prognostic relevance of the MC for different pathologists was only determined in one study.[126]

For all tumor types/groups combined, a prognostic value (mostly determined using statistical significance) was found in 55% to 63% of the studies regarding the different outcome metrics as summarized in Table 4. The association of the MC with survival for the tumor types/groups with 3 or more studies are summarized in Fig. 4. However, the discriminant ability of the prognostic test can not be properly evaluated for many studies due to the lack of appropriate statistical analysis.

Survival was evaluated in 23 studies on mast cell tumors of skin (cutaneous and subcutaneous). A shorter survival was found for cases with higher MCs in 21/23 (91%) studies, while 2/23 (9%) studies did not reach statistical significance. A relevant discriminant ability of the MC is suggested by the AUC values of 0.78, 0.79 and 0.82. The sensitivity and specificity values for the different proposed cut-off stratifications, reported in 9 individual studies, are illustrated in Fig. 5. Higher MCs were significantly associated with shorter disease-free intervals in 8/10 (80%) studies and with occurrence of metastasis or recurrence in each 4/6 studies (67%).

For cutaneous and oral melanoma, higher MC values indicated shorter survival 11/17 (65%) studies. Interestingly, analysis of only cutaneous melanoma found a prognostic significance in 4/5 studies (80%), analysis of only oral melanoma in 4/9 studies (44%) and combined analysis of both locations in 4/4 studies (100%). For cutaneous and oral melanoma, a good discriminant ability of the MC is inferred, based on the provided AUC values of 0.78, 0.79 and 0.86. Shorter disease progression was significantly associated with higher MCs in 3/5 studies (60%).

Studies on soft tissue sarcoma of the skin reached statistical significance regarding higher MCs and shorter survival in 6/7 studies (86%). The discriminant ability of the MC is not well demonstrated in these studies. Recurrence was associated with higher MCs in 7/10 studies (70%).

Studies on splenic stromal sarcoma and the former broader category "fibrohistiocytic nodules" found a significant prognostic association of higher MCs values with shorter survival in 3/4 (75%) instances. The failure to reach significance in the one study could be due to a small study population of 8 tumors.

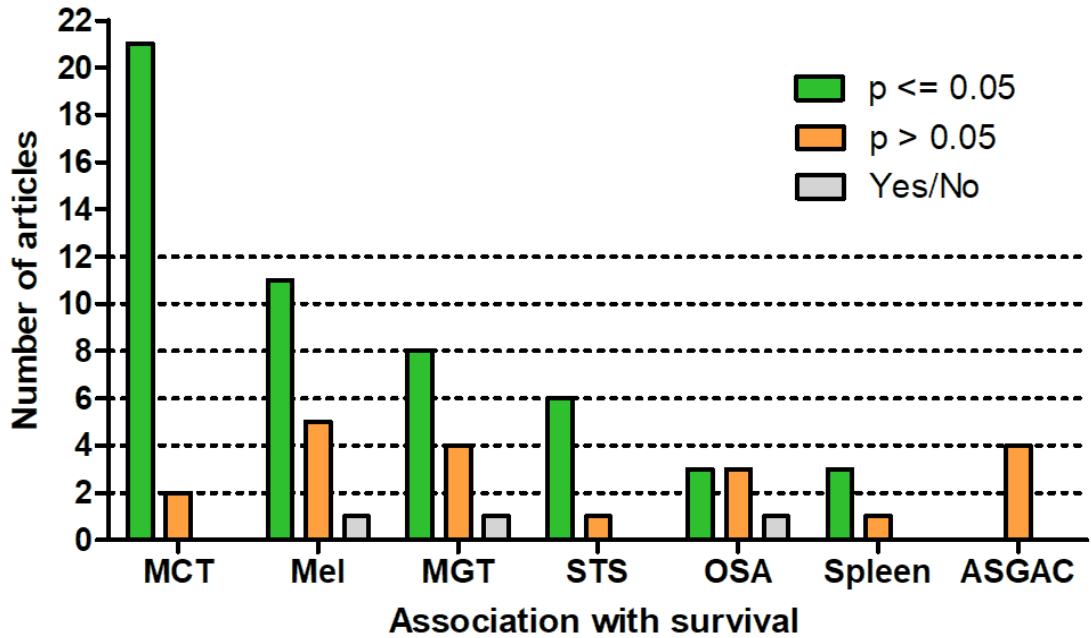

**Figure 4.** Number of studies that did or did not reach prognostic significance regarding the association of the MC with survival for tumor types with more than three studies. Yes/No describes studies with significant association in only a subset of cases or pathologists.

MCT, mast cell tumor of the skin (cutaneous and subcutaneous); Mel, cutaneous and oral melanoma; MGT, mammary gland tumors; STS, soft tissue sarcoma; OSA, osteosarcoma; Spleen, splenic stromal sarcoma and fibrohistiocytic nodules; ASGAC, anal sac gland adenocarcinoma;

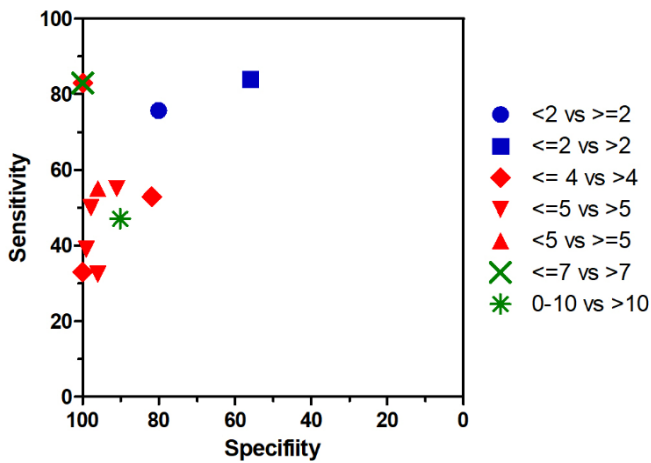

**Figure 5.** Summary of the sensitivity and specificity for survival in canine mast cell tumors of the skin based on different different cut-off ranges. The data is reported by 9 studies (three with two cut-offs).[10,11,19,51,55,60,145,151,153]

**Table 4.** Summary of the prognostic significance (mostly comprising the p-value approach) of all studies on the mitotic count combined regarding survival, disease progression (metastasis or recurrence), metastasis, and recurrence.

| Prognostic significance | Number of articles | | | | |
|---|---|---|---|---|---|
| | Survival | Disease progression | Metastasis | Tumor recurrence | Recurrence of clinical signs |
| Yes | 69 (63%) | 21 (55%) | 14 (58%) | 11 (55%) | 1 (100%) |
| Yes/No | 3 (3%) | 0 | 1 (4%) | 0 | 0 |
| No | 37 (34%) | 17 (45%) | 9 (38%) | 9 (45%) | 0 |
| ND | 4 | 2 | 2 | 2 | 1 |
| N/A | 19 | 92 | 106 | 110 | 130 |

Yes/No, prognostic significance was found for only a subset of the cases or pathologists; ND, interpretation of the prognostic significance is not provided (individual patient data available); N/A, outcome metric not available

### Mitotic index (MI): Risk of bias, MI methods, and prognostic value

The MI was determined in 5 studies that each evaluated a different tumor type: skeletal osteosarcoma,[95] multicentric lymphoma,[109] malignant mammary tumors,[124] mast cell tumors of the skin,[134] and aortic body tumors [159]. The overall risk of bias of these articles was judged to be high (n = 4) or moderate (n = 1) particularly pertaining to the data analysis domain.

The MI determination methods varied among studies. Four studies described the staining method, of which two used HE [109,159], one employed toluidine blue [124], and one utilized anti-PCNA immunolabeled slides.[134] The proportion of mitotic figures was calculated amongst varying numbers of tumor cells: 500,[109] 1,000 (in PCNA hotspot [134] or peripheral areas [95]), at least 10,000,[159] and all cells within 10 hotspot HPFs (at 25x magnification).[124] Two studies created photomicrographs for counting,[124,159] with one also using software for cell number estimation.[124]

None of the studies did reach statistical significance for the association of the MI with survival time (N = 2) or metastasis (N = 3). Only 1/2 studies determined that higher MIs were significantly associated with tumor recurrence.

## Discussion

The prognostic relevance of mitotic activity has been evaluated in many studies on canine tumors enabling this extensive systematic review. Canine studies on this topic were more numerous than feline studies, encompassing more than 3 times greater the numbers of articles,[14] and studies for other animal species are almost nonexistent.[6] This underscores an apparently greater research interest in canine tumors. Nevertheless, the findings of this systematic review on canine tumors was similar to the findings of a previous systematic review on feline tumors regarding the risk of bias in the studies.[14] For many tumor types the prognostic relevance of mitotic activity is still not convincingly proven considering the study limitations and lack of validation studies, as will be discussed below. Since our literature search for this systematic review, several articles on the MC have been published [21,29,37,49,64,73,74,82,120,148] and the number of new references is expected to markedly increase with time given the enormous increase of research interest over the last decade. Repetition of this systematic review will be required when new evidenced based literature allows new conclusions to be drawn on the prognostic value of the MC and MI in canine tumors.

### MC Methods

The MC represents the routine method of measuring mitotic activity with microscopic tumor evaluation. However, a wide variety of MC methods have been applied in previous studies and often the methods have not been described in sufficient detail.

Each study should describe the key aspects of the MC methods, including: 1) the region of interest selection, 2) area size (in mm²), and 3) spatial arrangement of HPF. While the best method regarding reproducibility and prognostic ability is current unknown, standardization, as previously proposed,[91] may improve comparability between studies and unify the diagnostic workflow between different tumor types. However, future studies are needed to determine which of these methods have the highest prognostic value and reproducibility between pathologists.[126]

Many laboratories have completely switched to digital microscopy [15] and high consistency with light microscopy has been shown by several studies (summarized by Donovan et al. [42]). Nevertheless, only few prognostic studies on canine tumors have used digital microscopy and correlated the digital counts with outcome.[154] Digital microscopy has some particular requirements such as the differences in size of HPFs that need to be considered for quantification of mitotic activity.[15,65] However, digital images also introduce new possibilities for standardized assessment of mitotic figures such as counting tools in viewing software and image analysis algorithms.[12,13] In particular, deep learning-based algorithms are considered promising to improve time efficiency, reproducibility and accuracy for this task (computer-assisted mitotic counts).[4,12]

### Prognostic Value of the MC

Tumor cell proliferation is one key driver of tumorigenesis, and thus the MC is often assumed to correlate with outcome. However, a surprisingly high proportion of studies did not find a prognostic value (using statistical significance) for the MC in canine tumors. It should be noted that interpretation of these results are difficult, since many studies restricted their analysis to tests of significance (p-value approach), which cannot be used to establish prognostic value (effect size) or lack thereof.[17] We recommend correlating the MC with relevant endpoints by multiple statistical methods including the Kaplan-Meier curves, hazard ratios, sensitivity and specificity. Cut-off agnostic methods like ROC curves and their area under the curves (AUC) are particularly preferred.[17]

Of note, conflicting findings between studies were found for most of the evaluated tumor types/groups. This highlights the general high risk of bias of observational studies and the need for several validation studies before sufficient evidence of the prognostic value of the MC can be guaranteed. Possible explanations for the lack of prognostic relevance in these studies include small study populations, heterogeneous tumor groups (different tumor entities or locations), variable MC methods, and, as will be discussed in more detail below, flawed statistical methods.

We have noted that the results for the prognostic relevance of the MC were quite variable between different studies (as demonstrated for mast cell tumors). Besides the aforementioned limitations, the differences in the results can be explained by variability between pathologists in assessing the MC. While a high degree of inconsistency between pathologists has been shown by several studies,[4,12,13,154] the influence on prognostication is largely unexplored.[126] For example, it has been determined that some pathologists are rather sensitive when distinguishing mitotic figures from imposters (resulting in higher MCs), while other pathologists are rather specific (resulting in lower MCs).[4,12] We argue that this variability between pathologists might have an important influence on the prognostically most meaningful cut-off values, possibly resulting in unexpected performance of the prognostic test when applied routinely by various pathologists in a diagnostic setting.

Based on our systematic review, we conclude that the MC has a prognostic value for canine mast cell tumors of the skin, cutaneous melanoma, soft tissue tumors of the skin and splenic stromal sarcoma, thus determination of the MC is recommended for routine diagnostic service for these tumor types. While the discriminant ability seems to be good for mast cell tumors of the skin and cutaneous melanoma, it requires further evaluation for the other tumor types. For anal sac gland adenocarcinoma, the MC truly seems to have little prognostic value based on several studies. The results for the prognostic value of the MC are conflicting for mammary tumors, oral melanoma and osteosarcoma and unproven for all the other tumor types considering the lack of validation studies.

## MI

In contrast to the MC, the MI has been rarely evaluated in the current literature, most likely explained by the inability to apply to routine diagnostic service. Improved time efficiency of the MI and thus applicability for routine diagnostics may be achieved in the future by the use of automated image analysis for tumor cells enumeration. It seems logical that the mitotic activity measurement is more representative for the case when set in relation to the cellular density, particularly in tumor types (such as mammary carcinoma) that exhibit variable cellular density due to extensive extracellular matrix, cystic spaces, inflammation or edema. Surprisingly, the few studies on canine tumors did not find a significant association with survival or metastasis, in contrast to the studies on feline mammary tumors.[14] However, the canine studies did not compare the MI to the MC and validation studies for each tumor type are not available.

## Conclusions

Mitotic activity is a relevant prognostic test that was evaluated in many studies on canine tumors. While the MI is rarely determined and its prognostic value is largely unexplored, the discriminant ability of the MC with regard to patient outcome has been well demonstrated in some canine tumors (particularly mast cell tumors of the skin, cutaneous melanoma and soft tissue tumors of the skin). Limitations of current studies include small case numbers, combined evaluation of heterogeneous tumor groups, unavailable detail of the MC methods, statistical analysis restricted to the p-value approach, and prognostic cut-offs based on single pathologists. Repetition of this systematic review is needed in several years to update conclusions and recommendations.

## Acknowledgements
None

## Declaration of conflict of interest
None

## Funding
None

## Authors' Contributions

CAB, TAD and AB designed the experiment; CAB performed the systematic review; CAB performed data extraction from the articles; the manuscript was written by CAB with contributions by all authors.